\documentclass[preprint,12pt]{elsarticle}

\usepackage{amssymb}
\usepackage{amsmath}

\usepackage{multirow}
\usepackage{booktabs}
\usepackage{hyperref}

\journal{Computer Physics Communications}

\begin{document}

\begin{frontmatter}

\title{ADAQ-SYM: Automated Symmetry Analysis of Defect Orbitals}

\author[inst1]{William Stenlund}

\affiliation[inst1]{organization={Department of Physics, Chemistry and Biology, Linköping University},
            city={Linköping},
            postcode={581 83}, 
            country={Sweden}}

\author[inst1]{Joel Davidsson}
\author[inst1]{Rickard Armiento}

\author[inst1,inst2,inst3]{Viktor Ivády}

\author[inst1]{Igor A. Abrikosov}

\affiliation[inst2]{organization={Department of Physics of Complex Systems, Eötvös Loránd University},
            addressline={Egyetem tér 1-3}, 
            city={Budapest},
            postcode={H-1053}, 
            country={Hungary}}
\affiliation[inst3]{organization={MTA–ELTE Lendület "Momentum" NewQubit Research Group},
            addressline={Pázmány Péter, Sétány 1/A}, 
            city={Budapest},
            postcode={1117}, 
            country={Hungary}}

\begin{abstract}
Quantum technologies like single photon emitters and qubits can be enabled by point defects in semiconductors, with the NV-center in diamond being the most prominent example. There are many different semiconductors, each potentially hosting interesting defects. 
The symmetry properties of the point defect orbitals can yield useful information about the behavior of the system, such as the interaction with polarized light.
We have developed a tool to perform symmetry analysis of point defect orbitals obtained by plane-wave density functional theory simulations. 
The software tool, named ADAQ-SYM, calculates the characters for each orbital, finds the irreducible representations, and uses selection rules to find which optical transitions are allowed. 
The capabilities of ADAQ-SYM are demonstrated on several defects in diamond and 4H-SiC.
The symmetry analysis explains the different zero phonon line (ZPL) polarization of the hk and kh divacancies in 4H-SiC.
\end{abstract}


\end{frontmatter}


{\bf PROGRAM SUMMARY}

\begin{small}
\noindent
{\em Program Title: ADAQ-SYM}                                          \\
{\em Developer's repository link:} https://github.com/WSten/ADAQ-SYM \\
{\em Licensing provisions:} GNU AFFERO GENERAL PUBLIC LICENSE Version 3  \\
{\em Programming language: Python 3}                                   \\
{\em Nature of problem:}\\
  Point defects in semiconductors can have localized orbitals in the band gap, these can be simulated with density functional theory (DFT). Automatically finding the symmetry properties (character and irreducible representation) of these orbitals would reduce manual work, and make the inclusion of symmetry properties in high-throughput screenings possible. 

\noindent
{\em Solution method:}\\
  ADAQ-SYM addresses this problem by calculating symmetry operator expectation values of orbitals computed with DFT, and translating these to characters and irreducible representation. The code also finds the symmetry allowed optical transitions.

\noindent
{\em Additional comments including restrictions and unusual features}\\
  Currently the code only works for DFT simulations at the $\Gamma$-point.
   \\

\end{small}

\section{Introduction}

Point defects in semiconductors can provide a platform for solid-state quantum technology, with applications such as qubits\cite{Koehl2011, doi:10.1063/5.0007444}, sensors \cite{PhysRevResearch.3.043007, Mzyk2022} and single photon emitters \cite{Berhane_2017, 10.1063/5.0006075}. One significant benefit of quantum applications made with solid-state point defects is room temperature operation \cite{Berhane_2017, doi:10.1063/5.0007444, Castelletto_2020}. 
Theoretical calculations have been proven useful for identification of potentially interesting defects in wide-band gap semiconductors and quantitative estimations of their properties \cite{Davidsson_mod_V_Si}.
Indeed, first-principles methods based on density functional theory (DFT) can simulate the electronic structures and predict multiple properties \cite{RevModPhys.86.253, Ivady2018, Dreyer}. 
Each semiconductor material may host a multitude of intrinsic and extrinsic point defects.
To probe the large combinatorically complex chemical space in an efficient manner, high-throughput workflows have been developed \cite{Sluydts2017, BROBERG2018165, DAVIDSSON_ADAQ} to simulate thousands of defect combinations, calculate relevant properties and store the results into a searchable database \cite{Sluydts2017, Bertoldo2022}.
Automatic Defect Analysis and Qualification (ADAQ) \cite{DAVIDSSON_ADAQ} is one such high-throughput workflow which currently has been used to screen about 52000 defects in 4H-SiC \cite{PhysRevB.108.224106} and 21000 defects in diamond \cite{davidsson2023na}.

Symmetry-based characterization is a well-established practice in analyzing point defect systems to understand the selection rules determining the polarization of absorbed and emitted light, to better understand features of the electronic structure such as degeneracies, and in general help understand and explain the physics of defects.
With symmetry analysis on the theoretical side, polarization specific PL measurements can be more accurately matched with simulated defects \cite{Davidsson_2020} and orientation for single defects can be identified \cite{PhysRevB.76.165205}.
Before analyzing the orbitals, the point group symmetry of the crystal hosting the defect needs to be found. There are two broadly used codes for this, spglib \cite{togo2018textttspglib} and AFLOW-SYM \cite{Hicks:AFLOW-SYM}. We use AFLOW-SYM because of its reported lowest mismatch when identifying space groups for known crystals from the Inorganic Crystal Structure Database (ICSD) \cite{Hicks:AFLOW-SYM}.
In addition, there are several codes that calculate irreducible representations of bands but mostly with focus on topological insulators \cite{GAO2021107760, IRAOLA2022108226, LIU2021107993, MATSUGATANI2021107948}.
Within quantum chemistry, one method to quantitatively analyse the symmetry of molecular orbitals is with continuous symmetry measures (CSM) \cite{doi:10.1021/ja00046a033, B911179D, C2CP41506B, Beevers2023}, which provide a numerical measure of how close molecular orbitals are to certain irreducible representations. 
Defect orbitals in the band gap are localized much like molecular orbitals, yet methods similar to CSM have not been applied to point defects in solid host materials. Presently, a common method of symmetry analysis of defect orbitals is visually inspecting an isosurface of the wave function and how it behaves under the symmetry transformations, this may be prone to human error especially for high symmetry structures, and is not applicable in high-throughput workflows. 
Another method of analyzing the symmetry is to describe the defect orbitals as a linear combination of atomic orbitals and performing a group theory analysis by making comparisons to basis functions of the character table \cite{PhysRevLett.101.226403, PhysRevMaterials.6.034601}. This method may be possible to automate, however, since the structure has already been relaxed with plane waves we focus on analyzing these directly without projecting to atomic orbitals. By omitting the projection we can keep all information present in the plane waves.  

This paper presents a quantitative symmetry analysis method for defect orbitals in solid host materials simulated with a plane wave basis set and the selection rules of optical transitions between defect orbitals. 

We introduce ADAQ-SYM, a Python implementation of this method. The tool is fast and automated, requiring little user input, making it applicable as an analysis tool for simulations of defects, also with applicability for high-throughput screening of defects.
Section~\ref{sec:theory} presents an introduction to the group theory, specifically applied to defects.
Section~\ref{sec:method} describes the ADAQ-SYM algorithm that performs the symmetry analysis, and Appendix~\ref{sec:imp} deals with how the software is constructed and what approximations are used. Computational details of the simulations in this paper are described in Section \ref{sec:comp details}.
Section~\ref{sec:results} presents the results from symmetry analyses of several known defects; nitrogen vacancy (NV) center and silicon vacancy (SiV) center in diamond and the silicon vacancy ($\mathrm{{V_{Si}}}$) and several divacancy ($\mathrm{V_{Si}V_C}$) configurations in 4H-SiC. Section \ref{sec:discussion} discusses these results and Appendix~\ref{sec:best practices} presents troubleshooting advice when using the tool. 

\section{Theoretical Background}
\label{sec:theory}

In this paper we consider point groups. For convenience we summarize basic concepts following Ref.~\cite{dresselhaus2008}.
We refer to symmetry transformations as unitary transformations in three dimensional space which have at least one fixed point, meaning no stretching or translation. In the Schönflies notation, these transformations are:

\begin{itemize} \label{sec:schoenflies}
    \item Identity, E.
    \item Rotation of \(2\pi/n\) or \(2\pi m/n\), where \(n\) and \(m\) are integers, \(C_n\) or \(C_n^{(m)}\).
    \item Reflection in a plane, \(\sigma_x\). \(x = h\), \(v\) or \(d\), denoting reflection in a horizontal, vertical or diagonal plane.
    \item Inversion, i.
    \item Improper rotation of \(2\pi/n\) or \(2\pi m/n\), where \(n\) and \(m\) are integers, which is a rotation \(2\pi/n\) or \(2\pi m/n\) followed by a reflection in a horizontal plane, \(S_n\) or \(S_n^{(m)}\).
\end{itemize} 
A set of these symmetry transformations, if they have a common fixed point and all leave the system or crystal structure invariant, constitutes the point group of that system or crystal structure.
The axis around which the rotation with the largest \(n\) occurs is called the principal axis of that point group.
The point groups relevant to solid materials are the 32 crystallographic point groups, of which the following four are used in this paper, $\mathrm{C_{1h}}$, $\mathrm{C_{2h}}$, $\mathrm{C_{3v}}$ and $\mathrm{D_{3d}}$.

In brief, character describes how a physical object transforms under a symmetry transformation, (1 = symmetric, -1 = anti-symmetric, 0 = orthogonal), and representation $\Gamma$ describes how an object transforms under the set of symmetry transformations in a point group.
Each point group has a character table which has classes of symmetry transformations on the columns and irreducible representation (IR) on the rows, with the entries in the table being characters. IRs can be seen as basis vectors for representations.
Each point group has an identity representation, which is an IR that is symmetric with respect to  all transformations of that point group.
Character tables can have additional columns with rotations and polynomial functions, showing which IR they transform as. 
Appendix \ref{sec:Character Tables} contains the character tables of the point groups used in this paper, these character tables also show how the linear (x, y and z) and quadratic ($\mathrm{x^{2}}$, xy, ...) polynomials transform.

\begin{figure}[h!]
\includegraphics[width=\columnwidth]{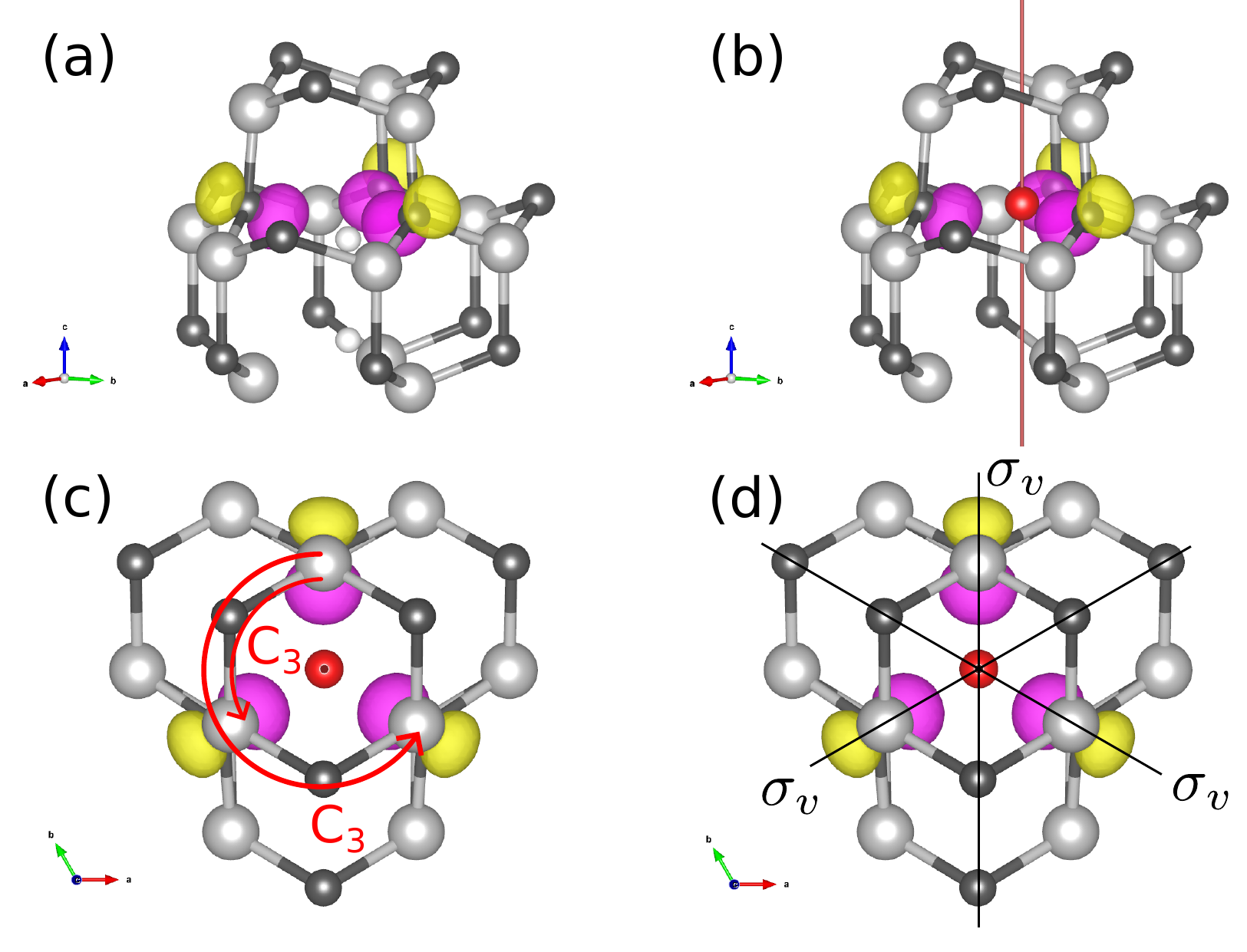}
\caption{Symmetries of a divacancy in 4H-SiC and one of its orbitals. Carbon atoms in black, silicon atoms in gray, positive and negative parts of the orbitals are represented by pink and yellow respectively. (a) Structure of the defect, vacancies are shown with small white spheres. (b) Center of mass of the orbital shown by red sphere, principal axis shown by red line. (c) Top view showing the two $C_3$ rotations. (d) Top view showing the three $\sigma_v$ reflection planes.}
\label{overlap showcase}
\end{figure}

For defects in solids, the point group is determined by the crystal structure, and the symmetry of the orbitals can be described by characters and IRs. Figure~\ref{overlap showcase} shows a divacancy defect in silicon carbide with the point group $\mathrm{C_{3v}}$ as an example.
Comparing with the character table for $\mathrm{C_{3v}}$, Table~\ref{table:Char table C3v} in Appendix \ref{sec:Character Tables}, one sees that the orbital has the IR $a_1$.

When defects are simulated with DFT, one obtains (one-electron) 
orbital wave functions $\phi_i$ and corresponding eigenvalues $\epsilon_i$. 
Optical transitions, where an electron moves from an initial state with orbital i to a final state with orbital f, has an associated transition dipole moment (TDM) $\vec{{\mu}}$ , which is expressed as:
\begin{equation}\label{TDM}
\vec{\mu} = \langle\phi_f|e\vec{r}|\phi_i\rangle, 
\end{equation}
where e is the electron charge and \(\vec{r}\) is the position operator.
Selection rules can be formulated with group theory \cite{dresselhaus2008}. For TDM the following applies:
for optical transitions to be allowed the representation of the TDM $\Gamma_{\mu}$ must contain the identity representation, where
\begin{equation}
\label{TDM selection rule}
\Gamma_{\mu} = \Gamma_f \otimes \Gamma_r \otimes \Gamma_i,
\end{equation} with $\otimes$ being the direct product, and $\Gamma_r$ is the IR of the polarization direction of the light, corresponding to the linear functions in the character tables.

\begin{figure*}[t]
\includegraphics[width=\textwidth]{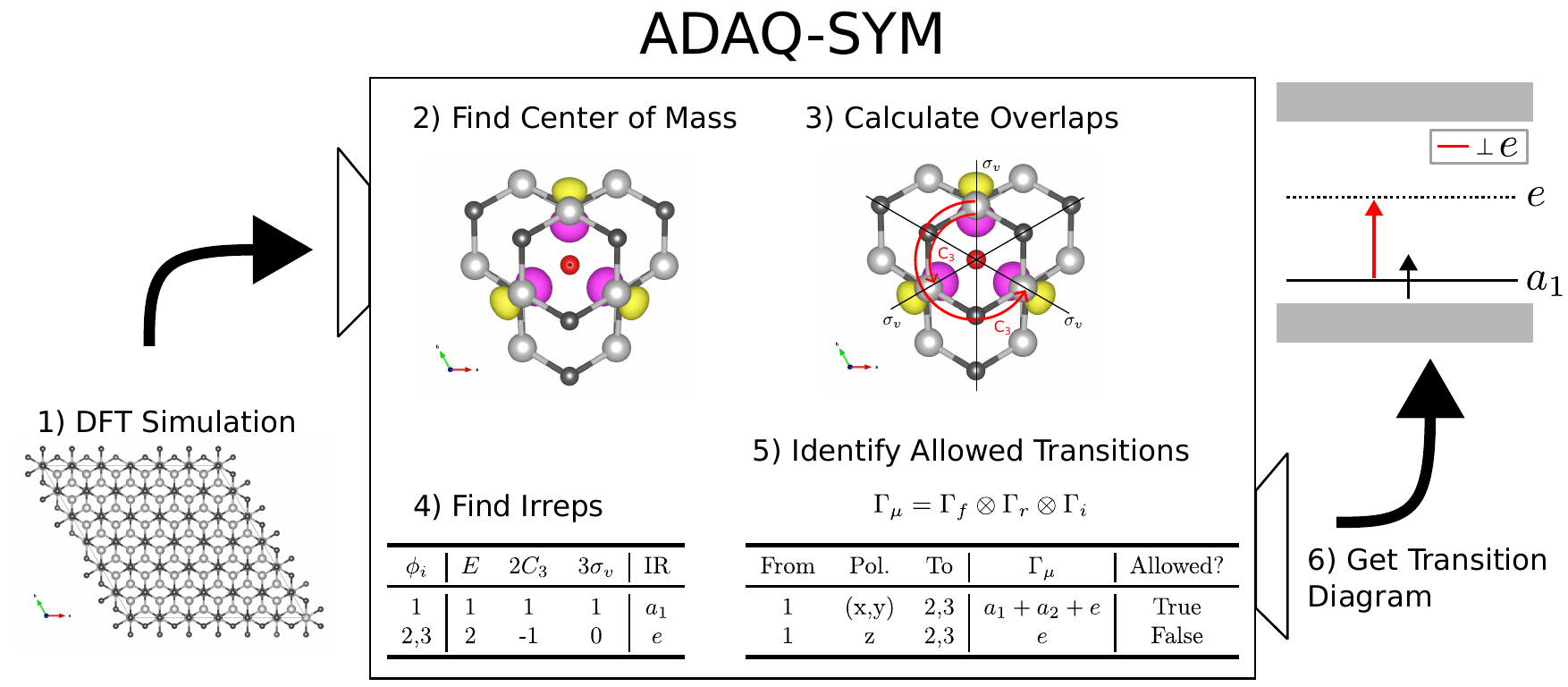}
\caption{ADAQ-SYM processes the output of a DFT simulation in steps, the center of mass is calculated for each orbital and is used as the fixed point for the overlap calculation. Then IRs are found from these overlaps and selection rules are applied for the transitions. Finally, a diagram of orbitals and transitions between them are generated.}
\label{symmetry analysis overview}
\end{figure*}

\section{Methodology}
\label{sec:method}
Figure~\ref{symmetry analysis overview} shows the symmetry analysis process of ADAQ-SYM. Here, we describe the steps in detail.

First, we perform a DFT simulation on a defect in a semiconductor host material. This produces a relaxed crystal structure and a set of orbital wave functions and their corresponding eigenvalues. These are the main inputs for ADAQ-SYM.
The electron orbitals associated with defects are localized around the defect, and the inverse participation ratio (IPR) $\chi$ is a good measure of how localized an orbital is \cite{IPR, IPR2, IPR3}. The discreet evaluation of IPR is 
\begin{equation} \label{IPR int}
    \chi = \frac{\sum\limits_r |\phi_i(\vec{r})|^4}{\Big(\sum\limits_r |\phi_i(\vec{r})|^2\Big)^2}, 
\end{equation}
and can be used to identify defect orbitals in the band gap, since they have much higher IPR than the bulk orbitals. There are also defect orbitals in the bands which are hybridized with the delocalized orbitals, their IPR are lower than the ones in the band gap, but still higher than the other orbitals in the bands. 
We employ IPR as a tool for selecting the orbitals to be analyzed, and this method can also identify defect orbitals in the bands by spotting outliers.
After the careful selection of ab initio data and inputs, ADAQ-SYM is able to perform the symmetry analysis.

Second, the "center of mass" $\vec{c}$ of the each orbital is calculated according to 
\begin{equation} \label{center of mass}
    \vec{c} = \langle\phi(\vec{r})|\vec{r}|\phi(\vec{r})\rangle = \int dr^3 \phi^*(\vec{r}) r \phi(\vec{r})) = \sum\limits_{r} \phi^*(\vec{r}) r \phi(\vec{r}).
\end{equation}
These centers are used as the fixed points for the symmetry transformations. Orbitals are considered degenerate if the difference in their eigenvalues are less than a threshold. When calculating $c$ for degenerate orbitals, they are considered together and the average center is used. 

This method does not consider periodic boundary conditions and necessitates the defect be in the middle of the unit cell. To mitigate skew of the center of mass, the wave function is sampled in real space, and points with moduli under a certain percentage $p$ of the maximum are set to zero according to
\begin{equation} \label{wf cutoff}
\phi_{trunc}(\vec{r}) = 
\begin{cases}
0 \: \: if  \: \: |\phi(\vec{r})| < p \max\limits_{\vec{r}}  (\phi(\vec{r}))  \\
\phi(\vec{r}) \: \: \text{otherwise} \\
\end{cases}
.
\end{equation}

Third, the point group and symmetry transformations of the crystal structure is found via existing codes. Each symmetry transformation has an operator $\hat{U}$. To get characters, the overlap of an orbital wave function and its symmetry transformed counterpart, the symmetry operator expectation value (SOEV), is calculated
\begin{equation} \label{ov integral}
    \langle\hat{U}\rangle = \langle\phi(\vec{r})|\hat{U}\phi(\vec{r})\rangle = \int dr^3 \phi^*(\vec{r}) (\hat{U}\phi(\vec{r})) ,
\end{equation}
for each orbital and symmetry transformation. The wave function is expanded in a plane wave basis set, with G-vectors within the energy cutoff radius. Therefore, Eq.~\ref{ov integral} can be rewritten to be evaluated by summing over these G-vectors only once, the plane wave expansion is also truncated by reducing the cutoff radius when reading the wave function and renormalizing \cite{masterthesis}.

Fourth, the character of a conjugacy class is taken to be the mean of the overlaps of operators within that class, and the overlaps of degenerate orbitals are added. 
To find the representation of a set of characters, the row of characters is projected on each IR, resulting in how many of each IR the representation contains. Consider an IR $\Gamma$, where \(\vec{W}_{\Gamma}\) is a vector with the characters of $\Gamma$ times their order. 
For example, for $\mathrm{C_{3v}}$ the vector for \(a_{2}\) is \( \vec{W}_{a_2} = (1\cdot1, 1\cdot2, -1\cdot3) = (1,2,-3)\). Let \(\vec{V}\) be the vector with a row of characters and \(h\) be the order of the point group, then \(N_{\Gamma}\) is the number of times the IR $\Gamma$ occurs which is calculated as follows 
\begin{equation}
\begin{split}\label{projection on IR}
N_{\Gamma} = \vec{W}_{\Gamma} \cdot \vec{V} \frac{1}{h} .
\end{split} 
\end{equation}
For degenerate states the found representation should be an IR with dimension equal to the degeneracy, e.g. double degenerate orbital should have a two-dimensional $e$ state. 
If an IR is not found, the overlap calculation is rerun with the center of another orbital as the fixed point.
The CSM $S$ for the IRs of molecular orbitals \cite{C2CP41506B} is used for the defect orbitals and calculated with 
\begin{equation} \label{csm}
    S(\phi, \Gamma) = 100(1-N_{\Gamma}), 
\end{equation}
which produces a number between 0 and 100. $S(\phi, \Gamma)=0$ means that the orbital is completely consistent with IR $\Gamma$, and $S(\phi, \Gamma)=100$ means that the orbital is completely inconsistent with the IR $\Gamma$.

Fifth, to calculate the IR of the TDM and find the allowed transitions, the characters of the TDM is calculated by taking the Hadamard (element-wise) product of the character vectors of each 'factor'
\begin{equation}
\begin{split}\label{TDM and character}
\vec{V}_{\mu} = \vec{V}_f \circ \vec{V}_r \circ \vec{V}_i \\
\end{split} 
\end{equation} and Eq.~\ref{projection on IR} is used to calculate $\Gamma_{\mu}$.
The representation of the resulting character vector is found in the same way as the IR of the orbital was found.
As an example, consider the group \(C_{3v}\) with the three IRs \(a_{1}\), \(a_{2}\) and \(e\). If some TDM in this group has the character vector \(\vec{V_{\mu}} = (4,1,0)\) calculating the representation would look like:
\begin{equation}
\begin{gathered}
\label{projection on IR example}
\vec{W}_{a_1} = (1,2,3), \: \vec{W}_{a_2} = (1,2,-3), \: \vec{W}_{e} = (2,-2,0), \: h = 6 ,\\
\Gamma_{\mu} = [N_{a_1},N_{a_2},N_{e}] = [\frac{4+2}{6}, \frac{4+2}{6}, \frac{8-2}{6}] = [1,1,1] .
\end{gathered}
\end{equation}
Since $\Gamma_{\mu}$ contains $a_{1}$, the transition is allowed. The software contains a function to convert a representation array of the above format to a string such as \("a_{1} + a_{2} + e"\).

Finally, the information produced by ADAQ-SYM is entered into a script to produce an energy level diagram which shows the position in the band gap, orbital occupation, IR and allowed transitions.

\section{Computational Details}
\label{sec:comp details}
The DFT simulations are executed with VASP \cite{VASP, VASP2}, using the projector augmented-wave method \cite{PAW, Kresse99}. 
We apply the periodic boundary conditions, and the defects in the adjacent supercells cause a degree of self-interaction. To limit this, the supercell needs to be sufficiently large. In our case supercells containing more than 500 atoms are used. 
The defects are simulated with two different exchange-correlation functionals, the semi-local functional by Perdew, Burke and Ernzerhof (PBE) \cite{PBE}, and the hybrid functional by Heyd, Scuseria, and Ernzerhof (HSE06)~\cite{HSE, HSEerrata}. The PBE simulations are presented in the supplementary material, and the HSE simulations are presented in section \ref{sec:results}.
These simulations only include the gamma point, run with a plane-wave cutoff energy of 600 eV, with the energy convergence parameters $1\times10^{-6}$ eV and $5\times10^{-5}$ eV for the electronic and ionic relaxations, respectively. The simulations are done without symmetry constraints so symmetry breaking due to the Jahn-Teller effect can occur when relaxing the crystal structure.
Excited states are simulated by constraining the electron occupation \cite{PhysRevLett.103.186404}. For cases where an orbital in the valance band is excited to a state in the band gap the setting \texttt{LDIAG=.FALSE.} is used.

\section{Results}
\label{sec:results}
To illustrate the capability of our method, we apply ADAQ-SYM to several defects in two different host materials, diamond and 4H-SiC, and analyze the symmetry properties for the defects orbitals.
The symmetry analysis provides a coherent picture of the known defects, finds the allowed optical transitions between defect orbitals, and, specifically, explains the different ZPL polarization of the hk and kh divacancies in 4H-SiC. The following results are from HSE simulations, see the supplementary materials for results in PBE, as well as experimental comparisons of band gaps and ZPL energies.

\subsection{Diamond Defects}

We first analyze the symmetry of the ground state of NV\textsuperscript{-}, SiV\textsuperscript{0} and SiV\textsuperscript{-} centers.
These defects were simulated in a cubic (4a,4a,4a) supercell containing 512 atoms, where $a=3.57$ Å.

\subsubsection{Negatively Charged NV Center}

\begin{figure}[h!]
\includegraphics[width=\columnwidth]{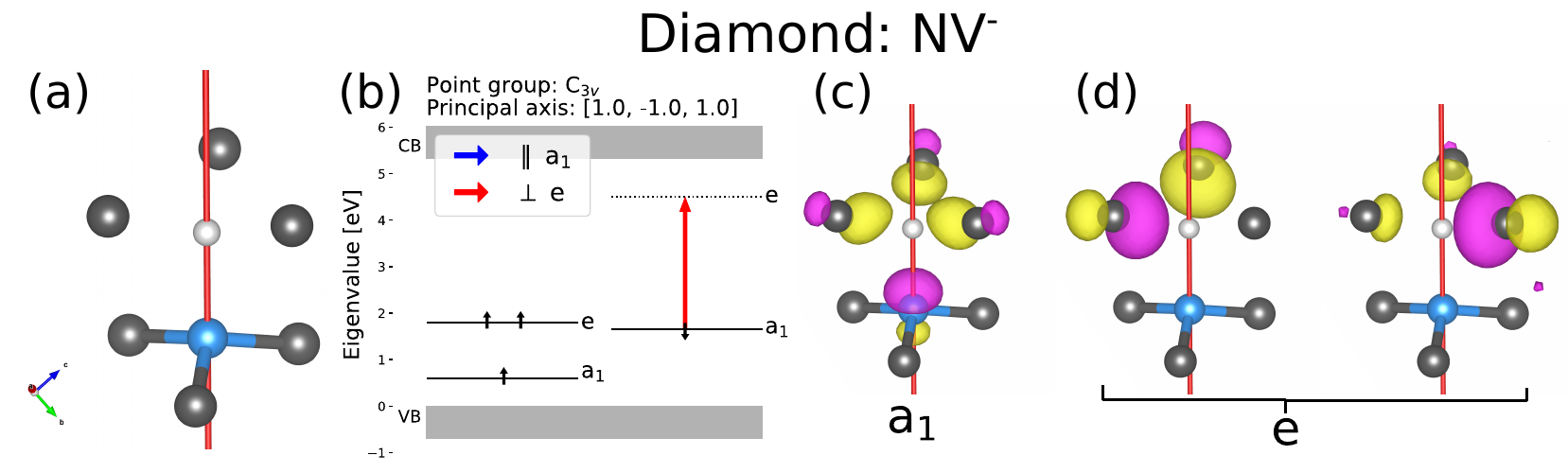}
\caption{Symmetry analysis of the negatively charged NV center in diamond. (a) The crystal structure of the NV\textsuperscript{-} center, where carbon atoms are black and nitrogen is blue, the vacancy is represented by a small white sphere. The principal axis (1,-1,1) is one of the body diagonals in the cubic unit cell, and is displayed as a red line. (b) Ground state Kohn-Sham orbitals and allowed transition. Eigenvalues are given in eV above the valance band maximum. Occupied spin down (up) states are marked with $\downarrow$ ($\uparrow$). Transitions are marked by arrows with color depending on the polarization of the absorbed (or emitted) light. Here only one transition is allowed, with polarization perpendicular to principal axis. The valance band (VB) and the conduction band (CB) are marked by gray rectangles. (c) The orbital of the occupied $a_1$ state. (d) The two orbitals constituting the empty degenerate $e$ state.  The positive and negative parts of the orbitals are represented by pink and yellow respectively.}
\label{transitions NV-}
\end{figure}

Figure \ref{transitions NV-} shows the ground state crystal structure and electronic structure of the NV\textsuperscript{-} center in diamond.
Figure \ref{transitions NV-} (b) is the generated output from ADAQ-SYM, for each orbital in the band gap. It shows the eigenvalue, occupation and IR, as well as the allowed transitions for each polarization. In this case, the found IRs are in accordance with previous work \cite{Doherty_2011}, and only one allowed transition is found, where the light is polarized perpendicular ($\perp$) to the principal axis. This selection rule has been experimentally confirmed \cite{PhysRevB.76.165205}.

\subsubsection{Silicon Vacancy Center}
\begin{figure}[h!]
\includegraphics[width=\columnwidth]{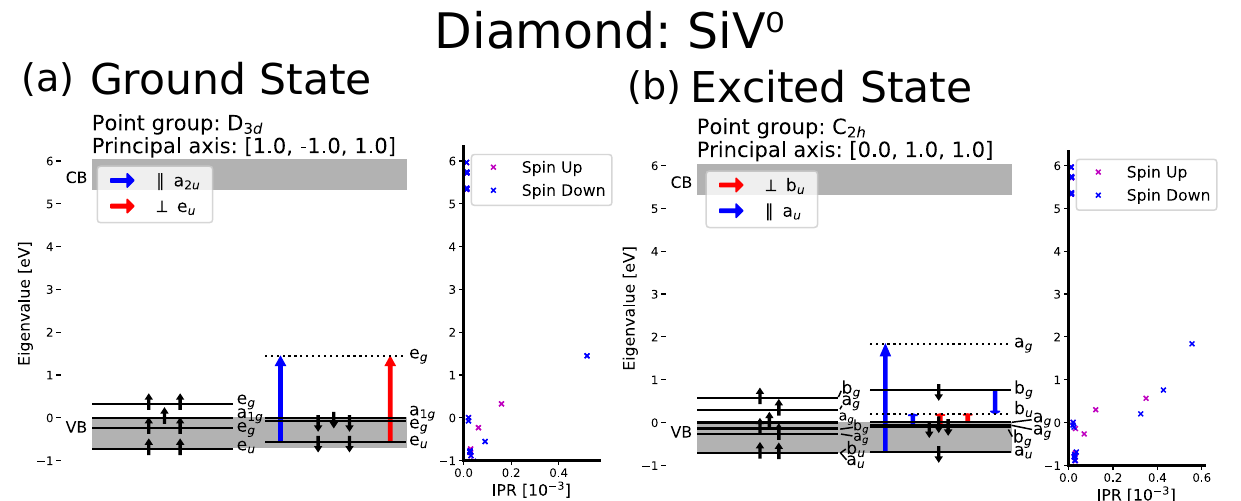}
\caption{Electronic structure of the neutral silicon vacancy center in diamond in the (a) ground and (b) excited state. IPR for 30 KS-orbitals around the band gap is shown right of each energy level diagram. KS-orbitals in the valence band are shown since there is a relevant transition between defect orbitals. Symmetry allowed transitions are marked with colored arrows denoting polarization. (a) Transitions from localized states in the valance band are symmetry allowed. (b) The unoccupied orbital is in the band gap, it is localized and ungerade. A Jahn-Teller distortion occurs in the excited state, lowering the point group symmetry from $D_{3d}$ to $C_{2h}$.}
\label{transitions SiV0}
\end{figure}

\begin{figure}[h!]
\includegraphics[width=\columnwidth]{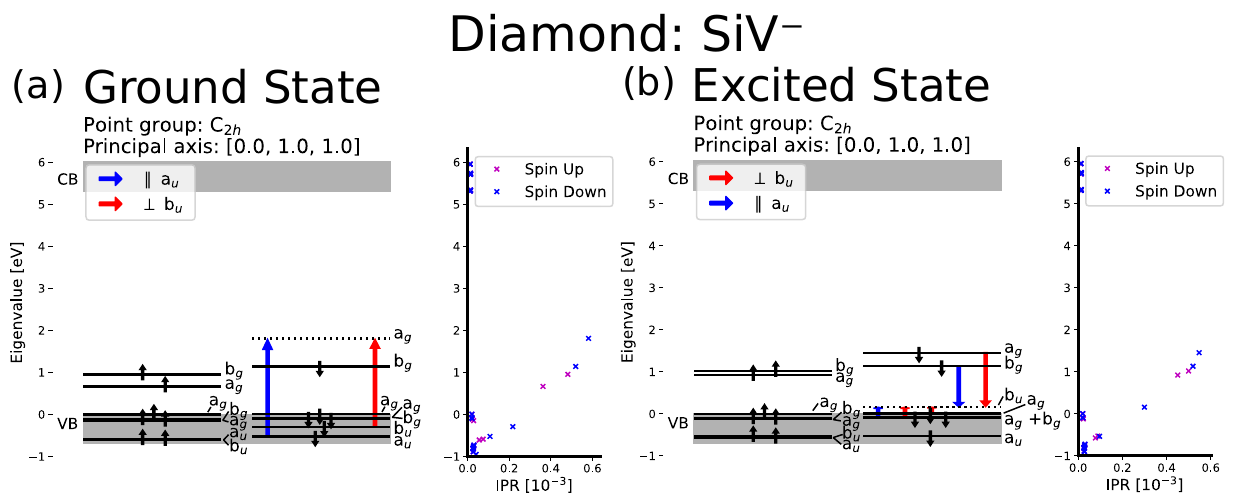}
\caption{Electronic structure of the negatively charged silicon vacancy center in diamond in the (a) ground and (b) exctied state. IPR for 30 KS-orbitals around the band gap is shown right of each energy level diagram. KS-orbitals in the valence band are shown since there is a relevant transition between defect orbitals. Symmetry allowed transitions are marked with colored arrows denoting polarization. (a) Transitions from localized states in the band gap are symmetry allowed. (b) The unoccupied orbital is in the band gap, it is localized and ungerade. Thus, there are symmetry allowed transitions between localized orbitals in the band gap.}
\label{transitions SiV-}
\end{figure}

Figures \ref{transitions SiV0} and \ref{transitions SiV-} show the electronic structure of the neutral and negatively charged silicon vacancy center in diamond, and the IPR for 30 KS-orbitals around the band gap. 
Our DFT calculations show that most orbitals in the VB are delocalized and have a low IPR.
However, some orbitals have larger IPRs meaning that they are more localized and indicating that they are defect states. These defect states in the VB are ungerade (u), meaning anti-symmetric with respect to inversion.
Both the charge states considered in this work have point groups with inversion symmetry which only allow optical transitions between orbitals of different symmetry with respect to inversion. To populate an orbital that is gerade (g), that is symmetric with respect to inversion, an electron from an u-state must be excited. When some orbitals in the valance band are taken into account, ADAQ-SYM finds two allowed transitions from localized defect states lower in the valance band to an empty state in the band gap, while transitions from the delocalized orbitals at the top of the valance band is forbidden. Both charge states has an empty orbital in the band gap in the excited state. For both $\mathrm{SiV^0}$ and $\mathrm{SiV^-}$ the localized state in the VB moves to the band gap in the excited state calculation, allowing for transitions from the other band gap states. These selection rules for the SiV defects are in agreement with previous calculations \cite{SiV, xiong2023midgap}. 

\subsection{Silicon Carbide}
In this subsection, we carry out the symmetry analysis of defects in 4H-SiC with ADAQ-SYM, in both the ground state and the lowest excited state. The IR of each KS-orbital in the band gap and the allowed polarization of light for both absorption and emission is shown in the figures below. 4H-SiC consists of alternating hexagonal ($h$) and quasi-cubic ($k$) layers, resulting in different defect configurations for the same stoichiometry. The defects were simulated in a hexagonal (6a,6a,2c) supercell containing 576 atoms, where $a=3.09$ Å and $c=10.12$ Å. For 4H-SiC, "in-plane" refers to the plane perpendicular to the c-axis.

\begin{figure}[h!]
\includegraphics[width=\columnwidth]{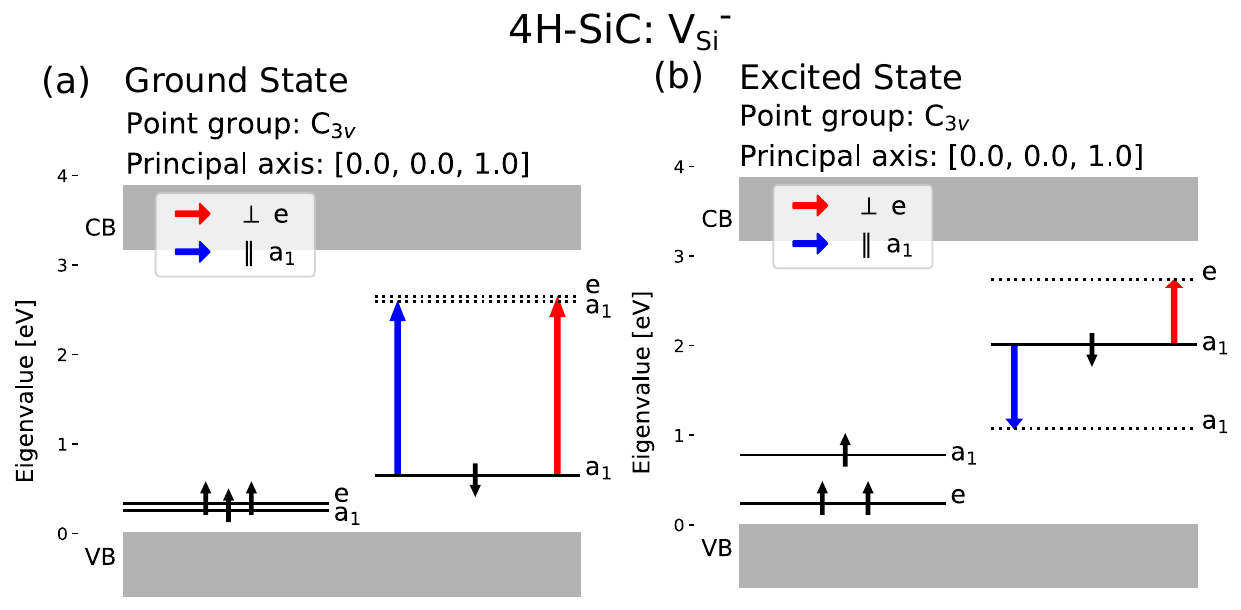}
\caption{Electronic structure of the (a) ground and (b) excited state of the silicon vacancy. In both states, the point group is $\mathrm{C_{3v}}$ and the principal axis is the c-axis. Allowed transitions are marked with colored arrows denoting polarization.}
\label{transitions V_Si}
\end{figure}

\subsubsection{Negatively Charged Silicon Vacancy}
We simulated the ground and excited state of the negatively charged silicon vacancy in the $h$ site.
Figure~\ref{transitions V_Si} (a) shows two allowed transitions with different polarization, where the parallel polarized transition has slightly lower energy than the perpendicular, this corresponds well to the V1 and V1' absorption lines \cite{JANZEN_V_Si} associated with the silicon vacancy in the h site \cite{PhysRevB.96.161114}.
Figure~\ref{transitions V_Si} (b) shows that the transition back to the ground state emits light polarized parallel to the c-axis, in agreement with previous calculations and measurements regarding the polarization of the V1 ZPL \cite{gali2012,Davidsson_mod_V_Si}.

\subsubsection{High Symmetry Divacancy}

\begin{figure}[h!]
\includegraphics[width=\columnwidth]{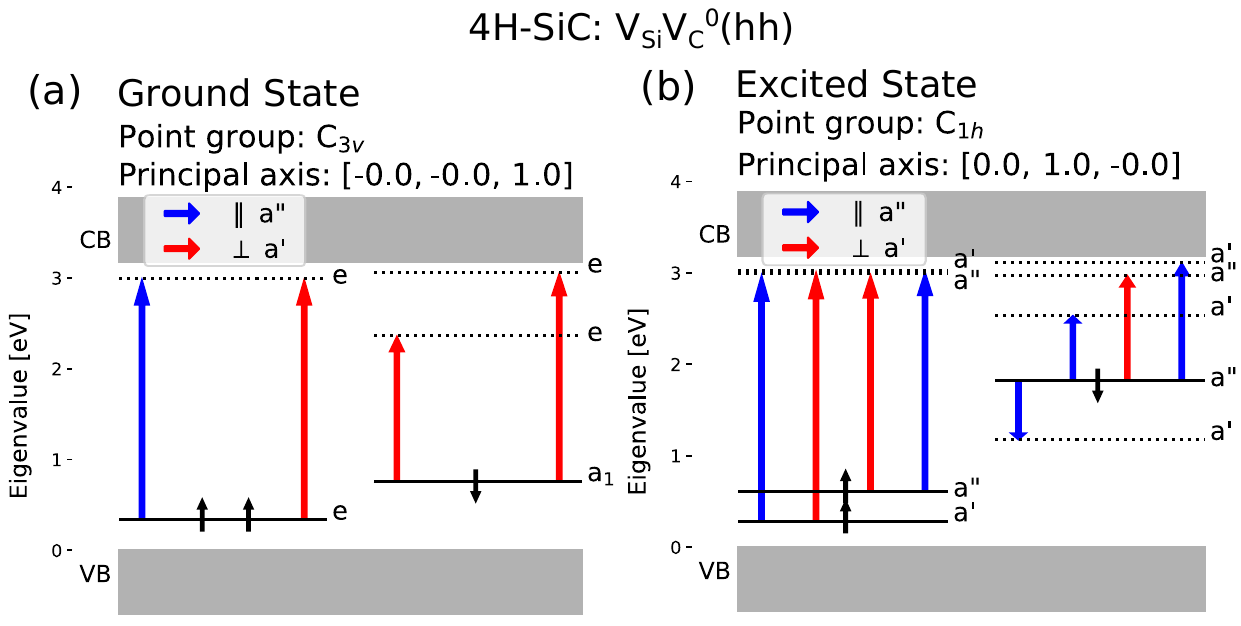}
\caption{Electronic structure of the neutral hh divacancy in the (a) ground and (b) excited state. In the ground state, the point group is $\mathrm{C_{3v}}$ with the c-axis as principal axis. In the excited state, the point group changes to $\mathrm{C_{1h}}$ with the principal axis lying in-plane. Note that the principal axis is given in the hexagonal lattice coordinates. Allowed transitions are marked with colored arrows denoting polarization.}
\label{transitions DiV hh}
\end{figure}

Figure \ref{transitions DiV hh} shows the ground and excited state of the $hh$ configuration of the divacancy, and the allowed transitions. In the excited state, one electron occupies what was previously an empty degenerate state and causes an Jahn-Teller effect. Because of this, the point group symmetry is reduced from $\mathrm{C_{3v}}$ to $\mathrm{C_{1h}}$ and degenerate states split when the system is relaxed in our simulations. This also changes the principal axis from being parallel to the c-axis to being perpendicular to it, that is the principal axis now lies in-plane. The selection rule tells us that absorption (to the lowest excited state) happens only for light polarized perpendicular to the c-axis, and the transition from the excited state emits light polarized parallel to the in-plane principal axis, thus also perpendicular to the c-axis. This polarization behavior corresponds well to previous calculations and measurements \cite{Davidsson_2020,Gallstrom812600}.
The $kk$ divacancy is  basically identical to the $hh$ divacancy with respect to symmetry.

\subsubsection{Distinguishing Low Symmetry Divacancies} 
\begin{figure}[h!]
\includegraphics[width=0.7\columnwidth]{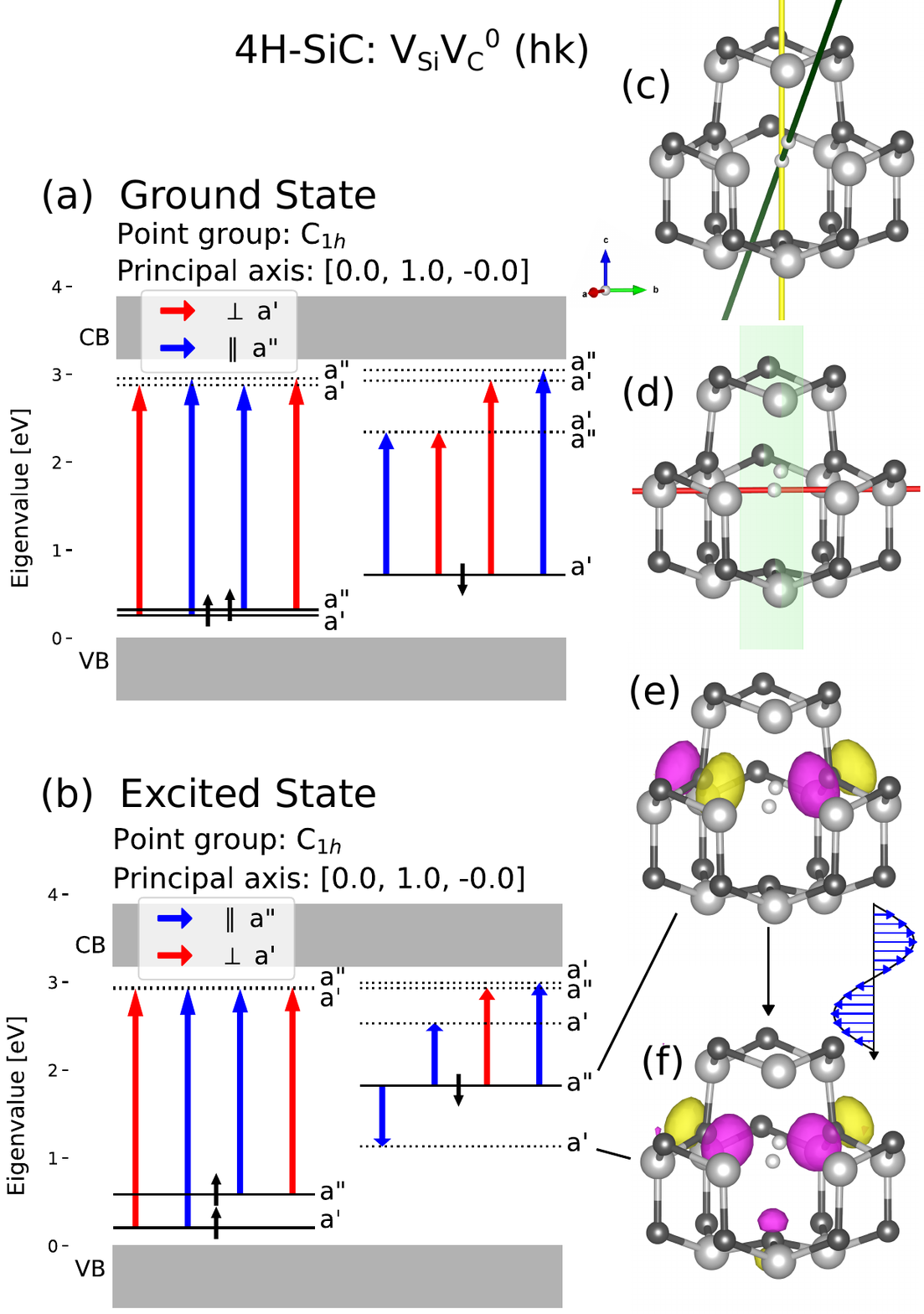}
\centering
\caption{Symmetry analysis for the neutral divacancy in 4H-SiC in the $hk$ configuration. (a)-(b) Electronic structure, ground state and excited state respectively. Allowed transitions are shown by colored arrows. (c) Crystal structure of the $hk$ divacancy with vacancies marked by small white spheres and the defect axis marked by the green line. The conventional c-axis is marked by the yellow line. (d) Principal axis represented by a red line and the perpendicular plane of reflection is shown in green. (e) Occupied anti-symmetric ($a''$) orbital in the excited state. (f) Occupied symmetric ($a'$) orbital in the ground state.}
\label{transitions DiV hk}
\end{figure}

\begin{figure}[h!]
\includegraphics[width=0.7\columnwidth]{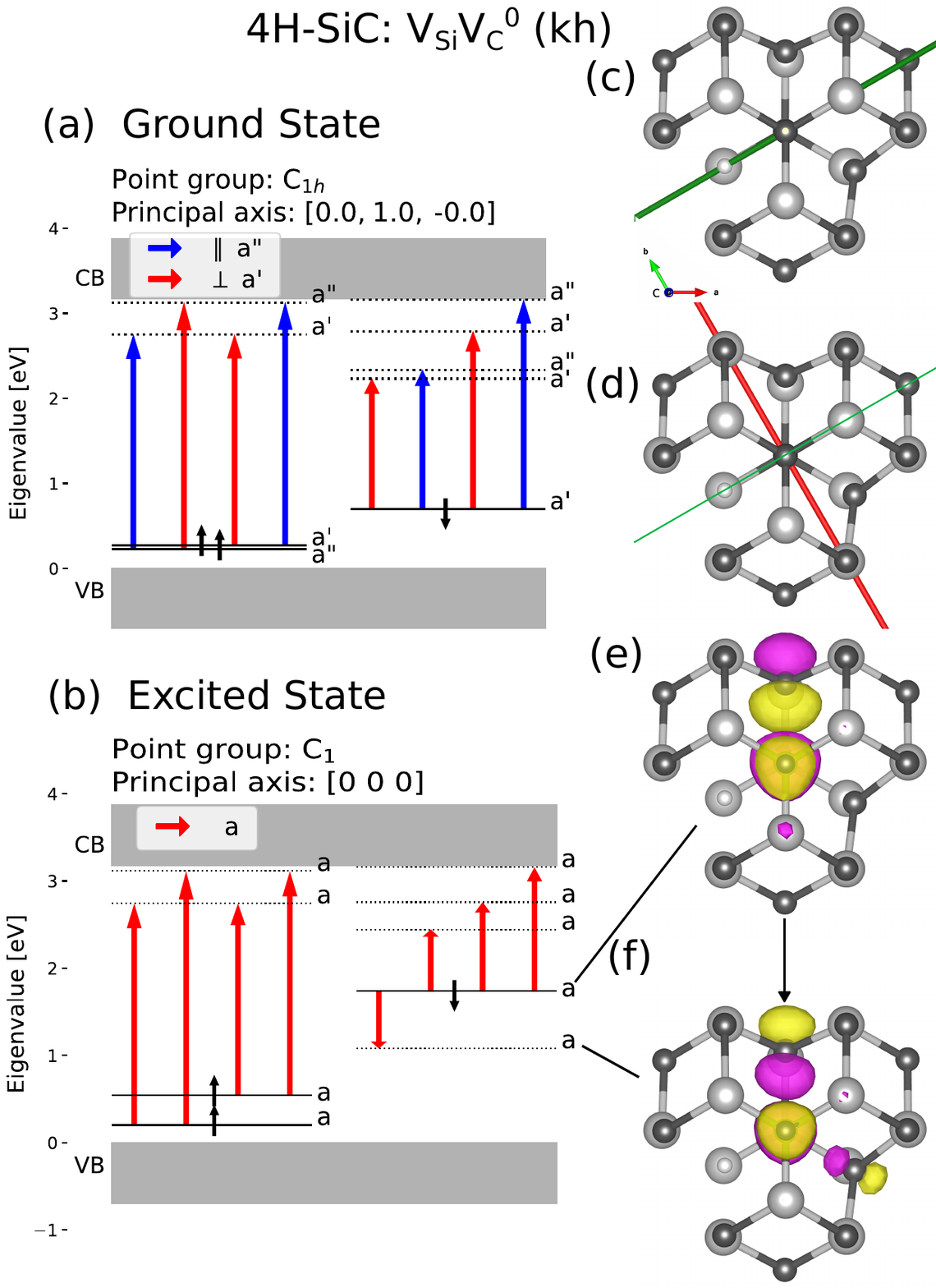}
\centering
\caption{Symmetry analysis for the neutral divacancy in 4H SiC in the $kh$ configuration. (a)-(b) Electronic structure, ground state and excited state respectively. Note that the symmetry has been reduced to $C_1$ in the excited state. Allowed transitions are shown by colored arrows. (c) Top view of the crystal structure of the $kh$ divacancy with vacancies marked by small white spheres (the second vacancy is behind the central C atom) and the defect axis marked by the green line. The conventional c-axis is point out of the figure. (d) Principal axis represented by a red line and the perpendicular plane of reflection is shown in green. (e) Occupied orbital in the excited state, not symmetric with respect to the point group of the crystal structure. (f) Occupied symmetric ($a'$) orbital in the ground state.}
\label{transitions DiV kh}
\end{figure}

The two low symmetry divacancy configurations $hk$ and $kh$ exhibit different behavior regarding the polarization of the ZPL \cite{Davidsson_2020,Gallstrom812600}. Examining the symmetry of the orbitals and applying selection rules regarding the TDM allows us to distinguish between these configurations. For both of these low symmetry configurations, the only symmetry transformation in the ground state is a reflection in a plane where the principal axis lies in-plane. 
Figure \ref{transitions DiV hk} shows crystal- and electronic structure information of the $hk$ divacancy. From panel (b) one sees that the relaxation to the ground state only emits light polarized parallel to the in-plane principal axis. 
Figure \ref{transitions DiV kh} shows crystal- and electronic structure information of the $kh$ divacancy. 
In the excited state the crystal is slightly symmetry broken from $C_{1h}$ to $C_{1}$, this is barely visible in panel (c). However, it is clear from the symmetry analysis of the relaxed ions. This symmetry breaking was not found in the PBE calculations and only appeared when the crystal structure for the excited state was relaxed using the HSE functional. Thus no optical transition is forbidden by the selection rules. Further analysis of the orbitals of the $kh$ excited state in the $C_{1h}$ point group is presented in the supplementary material, this analysis indicates the transition may be between two symmetric orbitals where the allowed polarization is perpendicular to an in-plane axis. In either case, excited state is C1h or C1, gives the same selection rules that are consistent with experimental results.

\section{Discussion}
\label{sec:discussion}

From the symmetry analysis by ADAQ-SYM, one can attribute the differing polarization behavior of the $hk$ and $kh$ configurations to the symmetry of the lowest excited state, anti-symmetric and asymmetric (possibly symmetric) respectively. Due to the principal axis laying in-plane, different defect orientations will also emit out-of-plane with different polarization. Thus, it may be possible to experimentally determine the orientation of individual defects measuring the in-plane polarization angle of the PL detected along the c-axis, in an experiment similar to Alegre et al. \cite{PhysRevB.76.165205}. In such an experiment, the $hk$ divacancy will exhibit a luminescence intensity maxima when the polarization is parallel to the principal axis, and a minima when the polarization in perpendicular. The opposite would be true for the $kh$ divacancy, and the two configurations could be distinguished by the approximately 30 meV difference in ZPL \cite{Davidsson_2020}, or by the 30 degree polarization differences between the respective maxima, see supplementary material for polar plots of this polarization dependence.

Having a loose tolerance parameter for the AFLOW-SYM crystal symmetry finder can be useful in ambiguous cases since ADAQ-SYM will then run for a larger set of symmetry operators, which gives an overview and can provide insight to what extent the orbitals are asymmetric with regards to each operator. It is also recommended to do this when multiple gradual distortions of the same defect are examined.
The initial PBE excited state calculation of the silicon vacancy in 4H-SiC seemed to show a case of the pseudo Jahn-Teller effect where the symmetry was reduced and the degenerate states split despite not being partially occupied in either spin channel. Upon running a simulation with more accurate tolerance parameters the point group remained $\mathrm{C_{3v}}$ and the splitting reduced to less than the threshold of 10 meV. This effect did not appear in the HSE simulation. For cases where the occupied state is very close to empty states convergence becomes more important and looser high-throughput simulations may exaggerate these effects. To resolve this one can either run more accurate calculations, or have a higher degeneracy tolerance parameter which will cause more states to be grouped together as degenerate. 

\section{Conclusion}
\label{sec:conclusion}
We have presented a method of determining the symmetry of defect orbitals, and implemented this method in the software ADAQ-SYM.
The implementation calculates the characters and irreducible representations of defect orbitals, the continuous symmetry measure is also calculated to get a numerical measure of how close the orbitals are described by the irreducible representations. Finally, ADAQ-SYM applies selection rules to the optical transitions between the orbitals. The tool is applicable to efficient analysis of defects.
We have applied the tool to a variety of known defects with different point groups and host materials, and it reliably reproduces their symmetry properties. 
In addition to the reproduction of prior results, we find that the polarization of the allowed transition for $hk$ is parallel to an in-plane axis, and while $kh$ exhibits slightly broken symmetry, the polarization is perpendicular to an in-plane axis. A method to determine the orientation of individual $hk$ and $kh$ divacancies is also proposed.
In summary, ADAQ-SYM is an automated defect symmetry analysis tool which is useful for both manual and high-throughput calculations.

\section*{Software Availability}
For availability of ADAQ-SYM and instructions, see \href{https://httk.org/adaq/}{https://httk.org/adaq/}.

\section*{Author contributions}
\textbf{William Stenlund}: Conceptualization, Investigation, Methodology, Software, Validation, Visualization, Writing - original draft.
\textbf{Joel Davidsson}: Conceptualization, Methodology, Supervision, Writing - review \& editing.
\textbf{Rickard Armiento}: Conceptualization, Supervision, Writing - review \& editing.
\textbf{Viktor Ivády}: Conceptualization, Methodology, Supervision, Writing - review \& editing.
\textbf{Igor A. Abrikosov}: Conceptualization, Supervision, Writing - review \& editing.

\section*{Declaration of Competing Interest}
The authors declare that they have no known competing financial interests or personal relationships that could have appeared to influence the work reported in this paper.

\section*{Acknowledgements}

This work was partially supported by the Knut and Alice Wallenberg Foundation through the Wallenberg Centre for Quantum Technology (WACQT).
We acknowledge support from the Knut and Alice Wallenberg Foundation (Grant No. 2018.0071).
Support from the Swedish Government Strategic Research Area Swedish e-science Research Centre (SeRC) and the Swedish Government Strategic Research Area in Materials Science on Functional Materials at Linköping University (Faculty Grant SFO-Mat-LiU No. 2009 00971) are gratefully acknowledged.
JD and RA acknowledge support from the Swedish Research Council (VR) Grant No. 2022-00276 and 2020-05402, respectively.
The computations were enabled by resources provided by the National Academic Infrastructure for Supercomputing in Sweden (NAISS) and the Swedish National Infrastructure for Computing (SNIC) at NSC, partially funded by the Swedish Research Council through grant agreements no. 2022-06725 and no. 2018-05973.
This research was  supported by the National Research, Development, and Innovation Office of Hungary  within the Quantum Information National Laboratory of Hungary (Grant No. 2022-2.1.1-NL-2022-00004) and within grant FK 145395.

\appendix

\section{Implementation}
\label{sec:imp}

ADAQ-SYM is written in python using functional programming. Table \ref{table:python functions} provides an overview of the principal functions, and Table \ref{table:settings} shows the settings ADAQ-SYM uses. To run the code, the user needs to provide three files from a VASP simulation; POSCAR or CONTCAR, the crystal structure; WAVECAR, wave function; EIGENVAL, eigenvalues and occupation of the bands. 
To select which bands get analyzed, either use the IPR threshold or select the bands manually.
For now, the code only works for defects simulated at the Gamma-point.

\begin{table}[h!]
\caption{Overview of the principal functions performing the symmetry analysis. 
The left column shows the name of the function. The right column shows the core output of each function. 
These functions may have more inputs and output than is described here, see the commented code for more detailed information. Various helper functions are excluded.}
\label{table:python functions}
\begin{tabular}{l|@{\hskip 2mm}l}
\toprule
Function name & Output\\
\midrule
\texttt{get\_symmetry\_operators()} & transformation matrices\\
\texttt{get\_point\_group()} & point group \\
\texttt{find\_average\_position()} & centers \\
\texttt{get\_overlaps\_of\_bands()} & overlaps \\
\texttt{get\_energy\_and\_band\_degen()} & degeneracy grouping\\
\texttt{get\_character()} & characters \\
\texttt{get\_rep()} & representation \\
\texttt{get\_allowed\_transitions()} & transitions \\
\texttt{plot\_levels()} & diagram \\
\bottomrule
\end{tabular}
\end{table}

\begin{table*}[t]
\small
\caption{Overview of the settings that are provided to ADAQ-SYM in a json-file. The left column gives the names of the setting. The center column provides a brief description of how the setting is used. The right column gives a recommended value of the setting.}
\label{table:settings}
\begin{tabular}{l@{\hskip 2mm}|@{\hskip 2mm}l@{\hskip 2mm}|@{\hskip 2mm}c}
\toprule
Name & Description & Default\\
\midrule
\texttt{aflow\_tolerance} & AFLOW-SYM tolerance parameter & tight\\ 
\texttt{degeneracy\_tolerance} & Max energy differance between degenerate bands & 0.01 \\
\texttt{IR\_tolerance} & Maximun deviation of $\mathrm{N_{\Gamma}}$ from integer & 0.05 \\
\texttt{Gvec\_reduction} & Cutoff energy of plane wave is multiplied by this & 0.15 \\
\texttt{realgrid\_mult} & Increases grid density when calculating \ref{center of mass} & 4 \\
\texttt{percent\_cutoff} & Percentage $p$ in Eq.~\ref{wf cutoff} & 0.40 \\
\bottomrule
\end{tabular}
\end{table*}

The functions \texttt{get\_point\_group()} and \texttt{get\_symmetry\_operators()} call AFLOW-SYM \cite{Hicks:AFLOW-SYM} to find the point group and symmetry operators of the input crystal structure, the symmetry operators are then sorted by their conjugacy class and arranged in the order the classes appear in the character table. These functions use the \texttt{aflow\_tolerance} setting which determine tolerance for asymmetry AFLOW-SYM uses, values may be "tight" or "loose". 
The point group is used to load the right character table from text files by Gernot Katzer \cite{char_tables}. The vaspwfc module in the VaspBandUnfolding package \cite{zheng_2019} is used for reading the WAVECAR file and working with the plane wave expansion of the wave function, and it also serves as the basis of the IPR calculations.

The \texttt{function find\_average\_position\_general()} calculates the "center of mass" of each of the considered bands using Eq.~\ref{center of mass} and \ref{wf cutoff}, where the cutoff percentage $p$ is read from the \texttt{percent\_cutoff} setting. The wave function is sampled in a real space grid where the \texttt{realgrid\_mult} setting makes the grid denser.
The function \texttt{get\_overlaps\_of\_bands()} loops through all considered orbitals and all symmetry operators and calculates the overlap, how Eq.~\ref{ov integral} is computed is described in more detail in \cite{masterthesis} and Numpy \cite{2020NumPy-Array} is used to accelerate the evaluation.
The evaluation time of the overlap calculation scales linearly with the number of G-vectors in the plane wave expansion. To speed up the software the series is truncated by multiplying the cutoff energy by the factor \texttt{Gvec\_reduction}. The cutoff energy corresponds to a radius in k-space and only G-vectors within the radius are used, so halving the cutoff energy gives roughly one eighth as many G-vectors. Truncating the series produces some error in the overlap, this error is relatively small for \texttt{Gvec\_reduction} larger than 0.1 \cite{masterthesis}, the symmetry does not depend strongly on the high frequency components of the plane wave expansion. Note that the overlap calculation will produce a complex number.
The function \texttt{get\_energy\_and\_band\_degen()} reads the EIGENVAL file and groups the considered bands by degeneracy. Two bands are considered degenerate if the difference in eigenvalue is less than \texttt{degeneracy\_tolerance}. This function also outputs the eigenvalue and occupation of the considered bands.

The function \texttt{get\_character()} takes the overlaps and the bands grouped by degeneracy and first adds the overlaps of degenerate bands for each symmetry operator, then the overlaps within each conjugacy class is averaged to produce the character. At this point, the character is complex valued but this is resolved with the following function.
The function \texttt{get\_rep()} takes a set of characters and computes Eq.~\ref{projection on IR} for all IRs of a point group, since the overlaps are in general complex $N_{\Gamma}$ will also be a complex number. Doing this for a truly symmetric orbital will produce a complex number with a small imaginary component and a real component close to an integer. For a set of characters to be said to transform as IR $\Gamma$, the imaginary component must be smaller than \texttt{IR\_tolerance} and the real component must be within \texttt{IR\_tolerance} of a non-zero integer. For example with a tolerance of 0.05, characters producing $N_{\Gamma}= 0.99 + 0.02i$ will be interpreted as transforming as IR $\Gamma$, while characters producing $N_{\Gamma}= 0.96 + 0.07i$ or $N_{\Gamma}= 0.92 + 0.03i$ will not. The same procedure is used when the CSM is calculated, since Eq.~\ref{csm} uses $N_{\Gamma}$.
The function \texttt{get\_allowed\_transitions()} calculates Eq.~\ref{TDM and character} for each occupied state i, each non-full state f and each linear function r. The representation is found with \texttt{get\_rep()} and if the trivial representation is contained, the transition is marked as allowed.
The function \texttt{plot\_levels()} uses Matplotlib \cite{Hunter:2007} to create energy level diagrams of the considered states with the occupation and IR drawn, and all allowed transitions represented by arrows between the bands. The color of the arrow differs on the polarization of the transition. The levels of the VB and CB are determined by the eigenvalues of the delocalized orbitals above and below the localized orbitals. For cases where defect orbitals are found in the VB or CB, e.g. when all localized orbitals are not adjacent in terms of eigenvalue, it is recommended set the VB and CB eigenvalues manually in plotting script, look for the eigenvalues of the delocalized orbitals at the VB and CB edges.

\section{Troubleshooting}
\label{sec:best practices}
The following summarizes our recommendations when running the tool:

\emph{If no IR is found and several bands are close, increase degeneracy tolerance which will cause more states to be grouped together as degenerate.} This may be preferential since actually degenerate orbitals split apart will not be assigned any IR, while accidentally degenerate orbitals grouped together as degenerate will assigned an IR which is the sum each orbitals IR, such as $\mathrm{a_{g}}+\mathrm{b_{g}}$, which makes it clear that the orbitals are accidentally degenerate.

\emph{If no IR is found, check that the centers of mass are close to your defect.} If not, recalculate the centers with higher grid density, setting \texttt{realgrid\_mult}  6 or 8. There is also an automated fallback where the atomic position of any unique atomic species will be used.

\emph{If the crystal symmetry is unclear or you think it should be higher, increase AFLOWs tolerance.} This way, the overlaps will be calculated for a larger set of symmetry operators.Then, check the overlaps manually, and look for subsets where the characters are close to integers, any such subset should be a point group which is a subset of the larger group.

\newpage
\section{Character tables}
\label{sec:Character Tables}
The character tables used in this paper are presented here. The tables include the linear and quadratic basis functions for the respective point groups.

\begin{table}[h]
  \caption{Character table of $\mathrm{C_{1h}}$.}
    \label{table:Char table C1h}
    \begin{tabular}{l|cc|ll}
     \toprule
      & E & $\sigma_h$ & Linear & Quadratic \\
     \midrule
     \(a'\) & 1 & 1 & x,y & $\mathrm{x^2, y^2, z^2, xy}$ \\
     \(a''\) & 1 & -1 & z & $\mathrm{xz, yz}$ \\
     \bottomrule
    \end{tabular}
\end{table}

\begin{table}[h]
  \caption{Character table of $\mathrm{C_{3v}}$.}
    \label{table:Char table C3v}
    \begin{tabular}{l|ccc|ll}
     \toprule
      & E & $2C_3$ & $3\sigma_v$ & Linear & Quadratic \\
     \midrule
     \(a_1\) & 1 & 1 & 1 & z & $\mathrm{x^2 + y^2, z^2}$ \\
     \(a_2\) & 1 & 1 & -1 & & \\
     \(e\) & 2 & -1 & 0 & (x,y) & $\mathrm{(x^2 - y^2, xy), (xz,yz)}$ \\
     \bottomrule
    \end{tabular}
\end{table}
\newpage
\begin{table}[h]
  \caption{Character table of $\mathrm{C_{2h}}$.}
    \label{table:Char table C2h}
    \begin{tabular}{l|cccc|ll}
     \toprule
      & E & $C_2$ & i & $\sigma_h$ & Linear & Quadratic \\
     \midrule
     \(a_g\) & 1 & 1 & 1 & 1 & & $\mathrm{x^2, y^2, z^2, xy}$ \\
     \(b_g\) & 1 & -1 & 1 & -1 & & $\mathrm{xz, yz}$ \\
     \(a_u\) & 1 & 1 & -1 & -1 & z & \\
     \(b_u\) & 1 & -1 & -1 & 1 & x,y & \\
     \bottomrule
    \end{tabular}
\end{table}

\begin{table}[h]
  \caption{Character table of $\mathrm{D_{3d}}$.}
    \label{table:Char table D3d}
    \begin{tabular}{l|cccccc|ll}
     \toprule
      & E & $2C_3$ & $3C'_2$ & i & $2S_6$ & $3\sigma_d$ & Linear & Quadratic \\
     \midrule
     \(a_{1g}\) & 1 & 1 & 1 & 1 & 1 & 1 & & $\mathrm{x^2 + y^2, z^2}$ \\
     \(a_{2g}\) & 1 & 1 & -1 & 1 & 1 & -1 & & \\
     \(e_g\) & 2 & -1 & 0 & 2 & -1 & 0 & & $\mathrm{(x^2 - y^2, xy), (xz,yz)}$ \\
     \(b_{1u}\) & 1 & 1 & 1 & -1 & -1 & -1 & &  \\
     \(b_{2u}\) & 1 & 1 & -1 & -1 & -1 & 1 & z & \\
     \(e_u\) & 2 & -1 & 0 & -2 & 1 & 0 & (x,y) & \\
     \bottomrule
    \end{tabular}
\end{table}

\bibliographystyle{elsarticle-num-names} 
\bibliography{refs}

\end{document}